\documentclass[pra,reprint]{revtex4-1}

\newcommand{\ket}[1]{|{#1}\rangle}
\newcommand{\bigket}[1]{\bigg|{#1}\bigg\rangle}
\newcommand{\bra}[1]{\langle{#1}|}

\newcommand{\be}{\begin{equation}}
\newcommand{\ee}{\end{equation}}

\usepackage{amsfonts}
\usepackage{amsmath}
\usepackage{rotating}
\usepackage{graphicx}
\usepackage{amssymb}
\usepackage{color}
\usepackage{epstopdf}
\usepackage{xfrac}

\begin{document}
\title{Mobile Spin Impurity in an Optical Lattice}

\author{C.~W. Duncan$^1$}
\email{cd130@hw.ac.uk}
\author{F.~F. Bellotti$^2$}
\author{P. \"Ohberg$^1$}
\author{N.~T. Zinner$^2$}
\author{M. Valiente$^1$}
\affiliation{$^1$SUPA, Institute of Photonics and Quantum Sciences, Heriot-Watt University, Edinburgh EH14 4AS, United Kingdom}
\affiliation{$^2$Department of Physics \& Astronomy, Aarhus University, Ny Munkegade 120, 8000 Aarhus C, Denmark}
\begin{abstract}
We investigate the Fermi polaron problem in a spin-1/2 Fermi gas in an optical lattice for the limit of both strong repulsive contact interactions and one dimension. In this limit, a polaronic-like behaviour is not expected, and the physics is that of a magnon or impurity. While the charge degrees of freedom of the system are frozen, the resulting tight-binding Hamiltonian for the impurity's spin exhibits an intriguing structure that strongly depends on the filling factor of the lattice potential. This filling dependency also transfers to the nature of the interactions for the case of two magnons and the important spin balanced case. At low filling, and up until near unit filling, the single impurity Hamiltonian faithfully reproduces a single-band, quasi-homogeneous tight-binding problem. As the filling is increased and the second band of the single particle spectrum of the periodic potential is progressively filled, the impurity Hamiltonian, at low energies, describes a single particle trapped in a multi-well potential. Interestingly, once the first two bands are fully filled, the impurity Hamiltonian is a near-perfect realisation of the Su-Schrieffer-Heeger model. Our studies, which go well beyond the single-band approximation, that is, the Hubbard model, pave the way for the realisation of interacting one-dimensional models of condensed matter physics.   
\end{abstract}

\maketitle

\section{Introduction}

Recently, strongly-interacting trapped one-dimensional multicomponent systems, which suffer from huge ground state degeneracies, have been shown to be tractable by means of freezing the charge degrees of freedom and the reduction of the spin sector to an effective spin chain model \cite{Volosniev2014,Deuretzbacher2014,Volosniev2015}. With this development, there has been considerable theoretical work on strongly interacting one-dimensional systems in recent years \cite{Cui2014,Dehkharghani2015,Yang2015,Massignan2015,Yang2016a,Yang2016b,Loft2016b,
Decamp2016,Hu2016b,Marchukov2016,Bellotti2016,Volosniev2016,Deuretzbacher2016b}, including for the case of a single spin impurity \cite{Levinsen2015,Loft2016c,Duncan2017}. As a result in the last year, numerical methods have been developed to obtain the effective spin chain from an arbitrary confining potential \cite{Loft2016a,Deuretzbacher2016a}. At the same time, ultracold atom experimental techniques have been developed to reach the few-body limit in one-dimensional set-ups \cite{Serwane2011,Endres2016}. There have been several experimental realisations of the few-body limit with fermions \cite{Zurn2012,Zurn2013,Wenz2013}, including for strong interactions \cite{Murmann2015}, and bosons \cite{EricTai2016}. 

The traditional notion of a polaron corresponds to a quasiparticle formed from the interactions between an impurity and its many-body surrounding medium, as first discussed by Landau and Pekar in 1948 \cite{Landau1948}. Polaron physics plays, for instance, an important role in the theory of superconductors with strong interactions, where the carriers are small lattice polarons and bipolarons \cite{Frohlich1954,Alexandrov2007}. There is also strong evidence that polarons play a role in the mechanism for some high-temperature superconductors \cite{Mott1993,Alexandrov2007,Devreese1996,*Devreese2000}. In magnetic systems, a spin polaron can be formed by the interaction of an impurity spin with the spins of the surrounding magnetic ions \cite{Devreese1996,*Devreese2000}.

It is well known that the definition of a quasiparticle becomes ill-defined in one dimension \cite{Kopp1990,Guan2013,Massignan2014}. The low-lying states for a single impurity fermion in one dimension in the homogeneous situation were derived by McGuire \cite{McGuire1965,McGuire1966}. The impurity problem in one dimension can also be considered in terms of a single, or two, particle-hole expansion, which gives a good approximation with fast convergence to the Bethe ansatz results \cite{Leskinen2010,Doggen2013,Astrakharchik2013}, and from which for attractive interactions a binding energy and effective mass of the impurity can be calculated \cite{Giraud2009}. The dressing of a single impurity fermion in one dimension by a majority Fermi sea has been considered experimentally \cite{Wenz2013}, providing a confirmation of the particle-hole expansion. This hints towards a polaronic-like behaviour for weak attractive interactions \cite{Guan2013,Massignan2014}, which has been studied theoretically, using the Fermi-Hubbard model, for the case of an imbalanced Fermi gas in an optical lattice \cite{Leskinen2010}. Evidence of polaronic behaviour of an impurity in a one-dimensional optical lattice has also been observed in the dynamics of a mobile spin impurity within the single-band Bose-Hubbard model \cite{Fukuhara2013}.

Polaron and impurity physics are also of great relevance in ultracold atomic physics. In this field, the polaron problem consists of a single impurity atom immersed in a many-body system of identical particles. The simplest problem of this kind corresponds to a fully polarised Fermi gas at very low temperature interacting with an atomic fermion of the same mass in a different hyperfine state. This is called the Fermi polaron problem \cite{Combescot2007,Prokofev2008} and has received considerable attention for almost a decade now \cite{Bruderer2007,Combescot2008,Leskinen2010,Chevy2010,Schmidt2012,
Doggen2014,Massignan2014,Yi2015}. In this time the Fermi polaron has been observed and investigated in several ultracold atom set-ups of different nature \cite{Schirotzek2009,Koschorreck2012,Chen2016,Scazza2016}. In addition, there have been experimental and theoretical works on the dynamics of an impurity \cite{Spethmann2012,Catani2012}, including a spin impurity in a one-dimensional lattice in the Hubbard model \cite{Massel2013,Fukuhara2013}.

Inspired by the capability of cold atom experiments, we consider the realisable scenario of a single spin impurity in a one-dimensional strongly repulsive Fermi gas in an optical lattice potential. This is the strongly repulsive one-dimensional limit of the Fermi polaron problem and goes beyond the single-band approximation of the Hubbard model. While the motivation for this work lies in the rich topic of polaronic physics, a polaronic-like behaviour is not expected in this limit. Throughout this work, we will refer to the state as that of a magnon or impurity.

In Sec.~\ref{sec:System}, we explicitly introduce the model we consider, including a brief discussion of the strongly interacting limit and the effective spin chain Hamiltonian of this limit. With the system defined we move on to discuss the dependence of the effective spin chain coefficients on the filling of the lattice in Sec.~\ref{sec:SpinCoefficients}. We will then consider the single impurity scenario in Sec.~\ref{sec:SinglePolaron}. In the final section, Sec.~\ref{sec:Interactions}, we extend the discussion to multiple magnons, with an emphasis on the nature of the interactions between them.

\section{System}
\label{sec:System}
We consider $N$ identical spin-1/2 fermions of mass $m$ in a one-dimensional periodic potential $V(x_i)$ with contact even-wave interactions of strength $g$. The Hamiltonian is then given by
\begin{equation}
H=\sum_i\left(\frac{p_i^2}{2m}  + V(x_i)   \right)+g\sum_{i<j}\delta(x_i-x_j).
\label{eq:Ham}
\end{equation}
The system is placed in a finite box of length $L$ with open boundary conditions. Throughout this paper, we set $\hbar = m = 1$, and express length in units of the length $L$ of the system. We consider the limit of strong repulsive interactions, $g \rightarrow +\infty$, for which the system can be mapped onto an effective spin chain model \cite{Volosniev2014,Deuretzbacher2014,Volosniev2015}. To make sure that the number of lattice wells is commensurate with the box's length, and without loss of generality, we choose a periodic potential of the form
\begin{equation}
V(x)=V_1\cos\left(\frac{2 \pi x}{d}\right) ,
\label{eq:Potential}
\end{equation}
where $d$ is the lattice spacing, defined as $d \equiv L/L_s$ with $L_s$ giving the number of wells -- or `sites' -- in the lattice. We consider a moderate lattice strength for all calculations of $V_1 = 5$. The filling factor $\nu$ of the lattice is defined as the number of particles per well of the optical lattice, i.e. $\nu \equiv N/L_s$. This will be the main parameter of the system.

\begin{figure}[t]
\begin{center}
\includegraphics[width=0.45\textwidth]{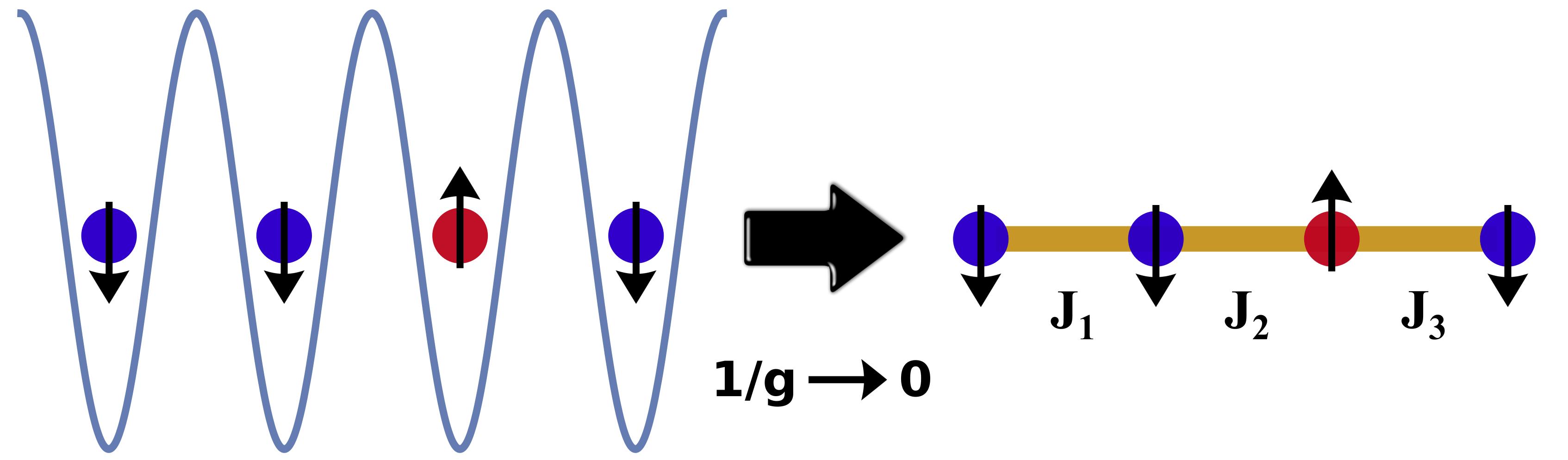}
\end{center}
\caption{Illustration of the mapping of the system to a spin chain with coupling constants $J_j$ when $g \rightarrow \infty$.}
\label{fig:Model}
\end{figure}

In the strongly interacting limit, $g \to \infty$, and at low energy, the charge degrees of freedom are fully described by $N$ spin-polarised non-interacting fermions. In this limit, to linear order in $1/g$, the dynamics of the spin degrees of freedom are described by an effective spin chain Hamiltonian \cite{Volosniev2014,Deuretzbacher2014,Volosniev2015}, illustrated in Fig.~\ref{fig:Model}. At $1/g \equiv 0$, the energy $E_0\equiv \lim_{g\to \infty}E(g)$ of the highly-degenerate ground state manifold is given by the spin-polarized fermionic, non-interacting ground state energy of Hamiltonian~(\ref{eq:Ham}). To order $1/g$, the energies in the ground state manifold are given by \cite{Volosniev2014,Cui2014}
\begin{equation}
E_n=E_0-\frac{K_n}{g},
\label{eq:KEnergy}
\end{equation}
for $n=1,\ldots,N_{\mathrm{deg}}$, where $N_{\mathrm{deg}}$ is the number of degenerate states in the manifold at $1/g=0$, and where $K_n$ ($>0$) is related to Tan's contact \cite{Olshanii2003,Tan2008,Braaten2008,Barth2011,Hofmann2012,
Werner2012,Valiente2011,Valiente2012a,Valiente2012b}, and is the $n$th eigenvalue of the effective spin chain Hamiltonian for the system
\begin{equation}
\hat{K} =-\frac{1}{2}\sum_{j=1}^{N-1}J_j\left(\boldsymbol{\sigma}_j\cdot \boldsymbol{\sigma}_{j+1}-1\right).
\label{eq:KHam}
\end{equation}
Above, $\boldsymbol{\sigma}_j=(\sigma_j^x,\sigma_j^y,\sigma_j^z)$ is the vector of spin-1/2 Pauli matrices operating at site $j$, and $J_j$ is the coupling coefficient between the $j$ and $j+1$ spins. Throughout this work, we will refer to $K_n$ as ``energies". As a result, the state with the highest $K_n$ corresponds to the ground state of the physical system for $g>0$.

The coupling constants $J_j$ depend exclusively on the trap's shape, strength and particle number \cite{Volosniev2014,Deuretzbacher2014,Volosniev2015,Loft2016a,Loft2016b}. Importantly, this is independent of the details of the spin degree of freedom. For atoms in optical lattices, the single particle solutions of the non-interacting system are Bloch waves, with $L_s$ states in each band. To calculate the spin chain coefficients we use the open source code CONAN \cite{Loft2016a}, which numerically calculates the coefficients for an arbitrary potential and up to $N \approx 35$ particles. From here on we set $N = 30$ unless otherwise stated, and scale through the filling of the lattice by varying the number of lattice wells $L_s$.

\section{Regimes of the System}
\label{sec:SpinCoefficients}

As stated above, the coefficients of the spin chain in the strongly-interacting limit are exclusively dependent on the single particle problem. Hence, the spin chain coefficients are solely dependent on the filling of the optical lattice. We consider the range of fillings $1/2 \leq \nu \leq 2$. Before calculating the coefficients of the spin chain using CONAN we can discuss their expected form, based on the fact that they are calculated from the $N$ lowest energy single particle states of the non-interacting part of Hamiltonian~(\ref{eq:Ham}). Over the range of filling considered the single particle spectrum goes from having a single band to two bands. For any $\nu$, the number of states in the first band can be written as $N - \theta(\nu - 1)(N-L_s)$ and the second band $\theta(\nu-1)(N-L_s)$, with $\theta$ denoting the Heaviside function.

\begin{figure}[t]
\begin{center}
\includegraphics[width=0.45\textwidth]{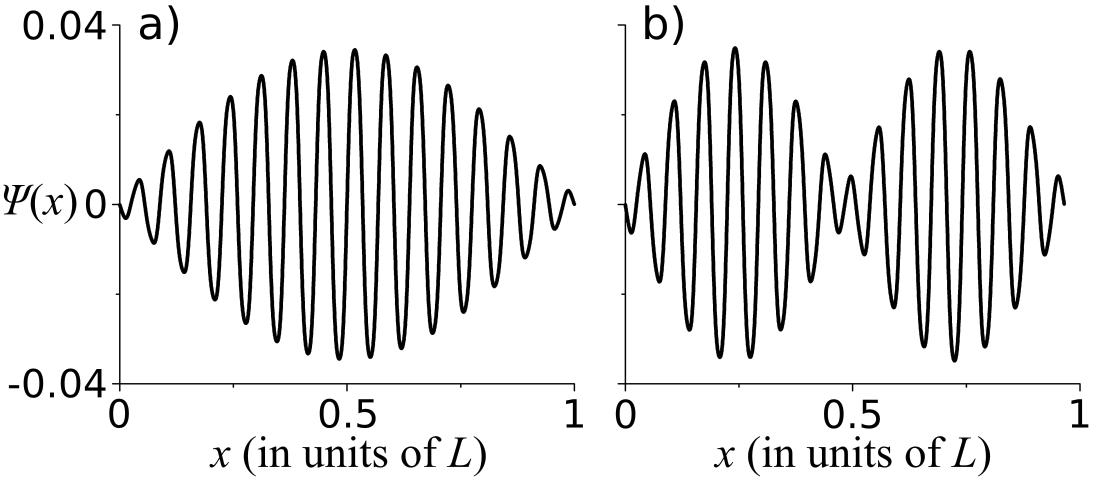}
\end{center}
\caption{Highest energy single particle eigenfunctions of the non-interacting part of Hamiltonian~(\ref{eq:Ham}) for a) $L_s = 29$ and b) $L_s = 28$ ($N = 30$).}
\label{fig:WavefunctionsSinglePart}
\end{figure}

In the next section, we will discuss the various regimes in the context of a single impurity or magnon. The system we consider has three distinct regimes:
\begin{enumerate}
\item $\nu \leq 1$; low filling case with particles occupying a single band. As a result, the spin chain coefficients will be homogeneous, with deviations only due to finite size effects.

\item $1 < \nu < 2$; high filling region with a two-band model of an unequal number of states in each band. In this regime, the spin chain coefficients are dominated by the $N-L_s$ states in the second band, which have a significant `box-like' component to the wavefunctions due to their high energy. This contribution from the box solutions defines the form of the coefficients, and, as we will discuss below, will result in the coefficients initially taking the form of a single (and multiple) inverted `well' potential.

\item $\nu = 2$; double filling is a special point, with two bands fully occupied. When mapped to the spin chain picture this filling results in a staggering of the spin chain coefficients between two values. This is analogous to the Su-Schriefer-Heeger (SSH) model \cite{Su1979,Su1980} of polyacetylene, and we compare the single magnon case to the SSH model in the next section.
\end{enumerate}

\begin{figure}[t]
\begin{center}
\includegraphics[width=0.5\textwidth]{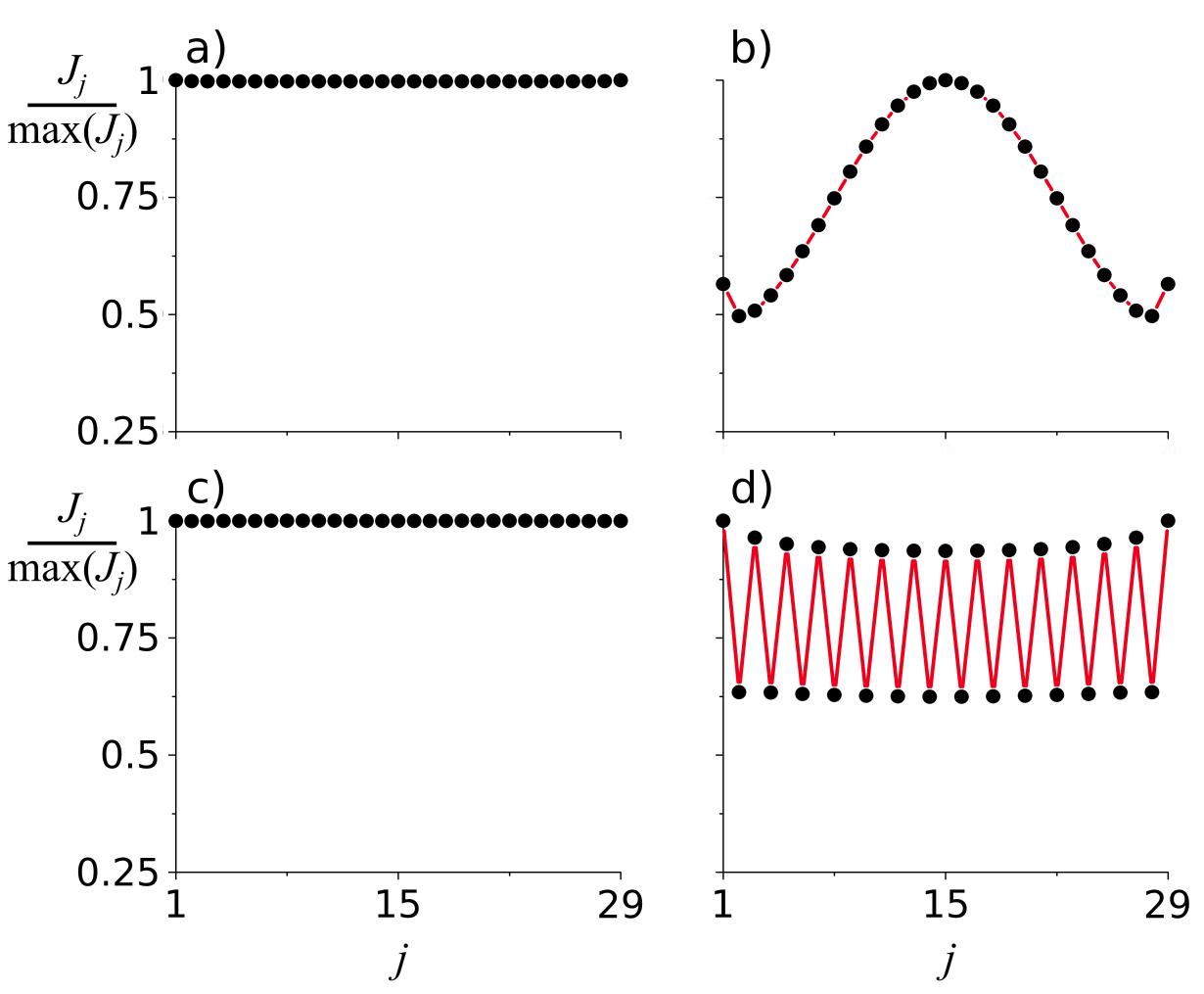}
\end{center}
\caption{Spin chain coefficients calculated by CONAN for $N = 30$ in the three distinct regimes of the system, circles (black) denote the values calculated and lines (red) are only included to help visualisation. The  coefficients for each regime have been normalized to their maximum. a) $\nu \leq 1$ case ($\nu = 1/2$ spin chain coefficients shown), b,c) $\nu > 1$ case (b) $p = 1$ and c) $p = 2$) and d) the $\nu = 2$ special point.}
\label{fig:CoefficientsExample}
\end{figure}

\begin{figure}[t]
\begin{center}
\includegraphics[width=0.5\textwidth]{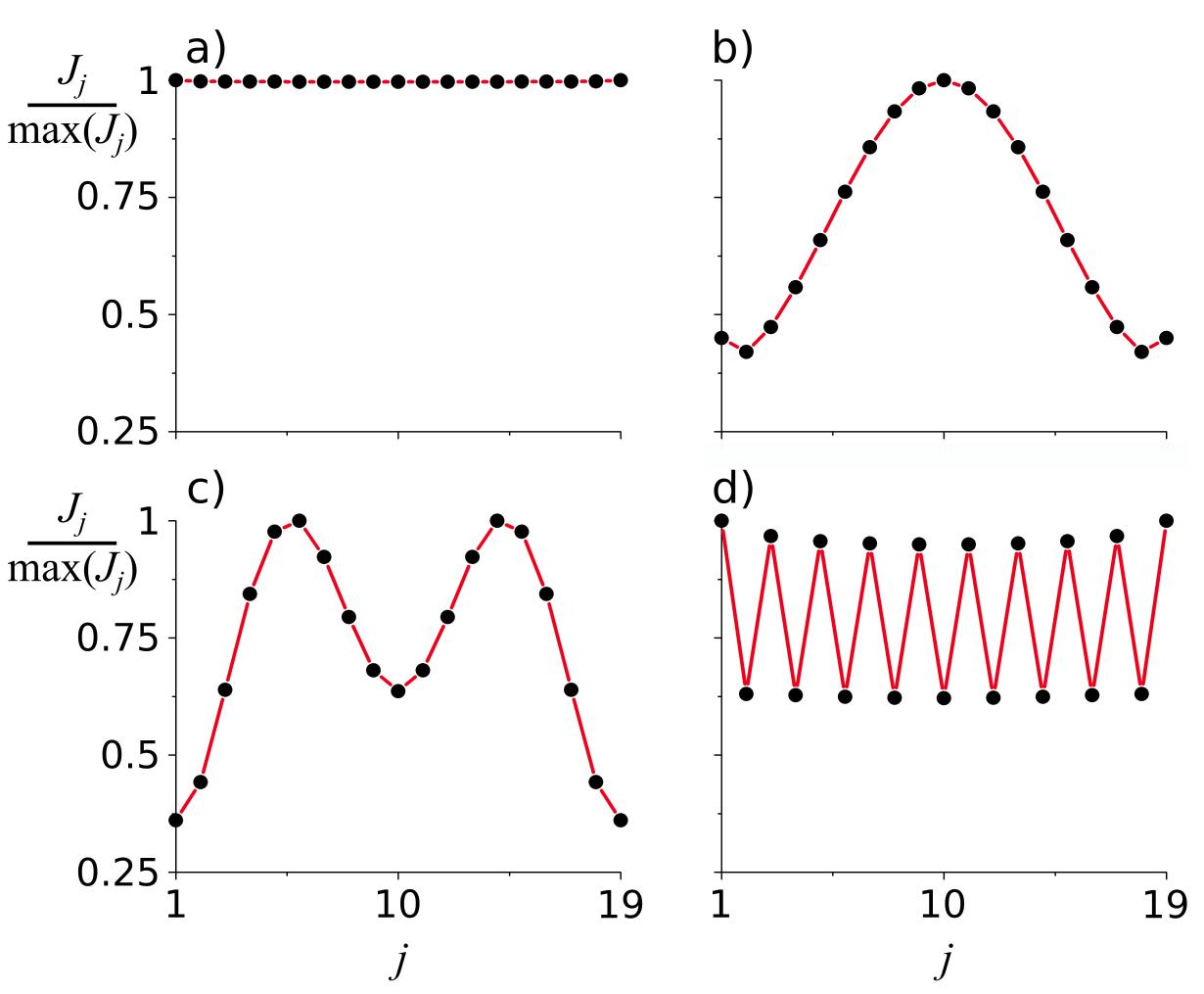}
\end{center}
\caption{Same as for Fig.~\ref{fig:CoefficientsExample}, except for $N=20$.}
\label{fig:CoefficientsExampleN20}
\end{figure}

As previously discussed, the spin chain coefficients depend exclusively on the single particle problem. In addition, the contribution of each single particle state is dependent, in part, on its energy (see Ref. \cite{Loft2016a} and references therein). As a result, for the $\nu > 1$ case, the largest contributions to the coefficients come from the states in the second band. For a filling of $\nu=1+p/L_s$, with $p \leq L_s$ and $p \in \mathbb{Z}$, $p$ states are occupied in the second band. The $p$ states of the second band are largely influenced by the hard wall boundary, due to their large energy and the moderate strength of the optical lattice, as can be seen in Fig.~\ref{fig:WavefunctionsSinglePart}. For a filling corresponding to $p = 1$, the single state in the second band, see Fig.~\ref{fig:WavefunctionsSinglePart}a, heavily dominates the form of the spin chain coefficients, which take the form of an inverted well, see Fig.~\ref{fig:CoefficientsExample}b. For $p = 2$ the coefficients have an inverted double well form, due to the highest energy state which is shown in Fig.~\ref{fig:WavefunctionsSinglePart}b. However, the well height for this filling is small, on the scale of $\sim 6 \times 10^{-4} \: \mathrm{ max}(J_j)$, which is effectively homogeneous, recovering the first regime, as seen in Fig.~\ref{fig:CoefficientsExample}c. The multiple well regime of the spin chain lasts, on a meaningful scale, until $p \sim N/4$, after which there is an extended crossover region to the third regime of staggered coefficients. The transition from the `well-like' structures to the special point of $\nu = 2$, while distinct, is extended over the region of filling approaching $\nu = 2$, for all $N$.

As discussed above, in the regime of $\nu > 1$ the form of the spin chain coefficients is a result of the hard-wall boundaries of the system. In Fig.~\ref{fig:CoefficientsExample}c we observe that the inverted `well-like' regime crosses over to an effectively homogeneous form quickly. This is entirely a result of the system size being rather large in one dimension, with finite-size effects only being substantial in the special case of a single atom over unit filling. Reducing the atom number will allow for multiple wells in the coefficients to be observed on a larger scale, e.g. for $N = 20$ as shown in Fig.~\ref{fig:CoefficientsExampleN20} (this case is considered in more detail in Appendix.~\ref{App:Coefficients}). When we move on to discuss multiple magnons in Sec.~\ref{sec:Interactions}, the extended regime of multiple inverted `wells' for $N \sim 20$ will be important.

At this point, it is worth noting that the choice of lattice strength will affect the numerical values obtained throughout this work. However, the validity of the three regimes, which are the main focus of this paper, is away from the extreme limits of the potential strength, i.e. for a moderate lattice strength. In the limit of a strong lattice strength, the $\nu<1$ homogeneous regime will still be present, however, the $\nu>1$ regime will change in nature as the excited states of the single particle spectrum will no longer be largely influenced by the hard wall boundary. The presence of the $\nu = 2$ special point does not depend on the lattice strength. In the weak lattice strength limit, the spin chain coefficients tend towards being homogeneous for all $\nu$, as is the case for no potential.

\section{Single Impurity}
\label{sec:SinglePolaron}

We consider a single spin-down fermion -- the impurity -- interacting with $N-1$ spin-up fermions. In the strongly-interacting limit, the spin of the impurity, in the spin chain, represents a single magnon. We have a basis of $N$ possible states e.g., for $N=4$, we have the basis
\begin{align*}
\mid \downarrow \uparrow \uparrow \uparrow \rangle \mbox{, }
\mid \uparrow \downarrow \uparrow \uparrow \rangle \mbox{, }
\mid \uparrow \uparrow \downarrow \uparrow \rangle \mbox{, }
\mid \uparrow \uparrow \uparrow \downarrow \rangle.
\end{align*}
For convenience, we will denote each state as $\ket{j}$, with $j$ giving the position in the spin chain of the spin-down fermion, allowing the wavefunction to be written as $\ket{\Psi} = \sum_j \psi (j) \ket{j}$.

There is a simple mapping of the single spin-down fermion spin chain Hamiltonian to a single particle tight-binding model with analogous hopping and potential. We can write the analogous single particle Hamiltonian as
\begin{equation}
\hat{K}^{(1)}= \sum_{j=1}^{N} \left[ t_j \left( \hat{b}^{\dagger}_j \hat{b}_{j+1} + \hat{b}^{\dagger}_{j+1} \hat{b}_{j} \right) + U_j \hat{b}^{\dagger}_j \hat{b}_j \right],
\label{eq:ParticleHam}
\end{equation}
where $t_j$ gives the the analogous hopping coefficient, $U_j$ the analogous on-site potential and $\hat{b}_j^{\dagger}=\ket{j}\bra{\mathrm{vac}}$ ($\hat{b}_j=(\hat{b}_j^{\dagger})^{\dagger}$) are single particle creation (annihilation) operators, with $\ket{\mathrm{vac}}$ the normalised vacuum state. The relations from the analogous parameters to the spin chain coefficients are
\begin{align}
t_j = & \: - J_j \label{eq:MappingTunnelingTightBinding} \\
U_j = & \: J_{j-1} + J_j \label{eq:MappingPotnentialTightBinding},
\end{align}
further details of the mapping are given in Appendix~\ref{app:TightBinding}. Note, $j$ is an index of the spin chain sites - the original fermions - and has nothing to do with the wells (or sites) of the lattice potential, with $j = 1,2,\dots, N$.

\subsection{Low Filling Regime}
\label{sec:Low}

For $\nu \leq 1$, the spin chain has a set of homogeneous coefficients, observed in Fig.~\ref{fig:CoefficientsExample}a. For the magnon/single particle analogy the hopping and on-site potential are constant for $j\neq 1,N$, i.e. $t_j = t$, $U_j = U$, with $t \approx - 2 U$, and $t,U > 0$. At sites $j=1,N$ the on-site potential has a value of $U_{1,N} = U/2$ due to the hard-wall boundaries. The effect of this inhomogeneity in Hamiltonian~(\ref{eq:ParticleHam}) is, remarkably, to make the magnon behave more as if the spin chain had periodic boundary conditions in the highest energy state of the system (lowest $K_n$), as we will show below.

We solve the stationary Schr\"{o}dinger equation for Hamiltonian~(\ref{eq:ParticleHam}) with the potential and hopping as specified above (for a detailed derivation see Appendix~\ref{App:LowFilling}). We find the $n$th eigenfunction in the spectrum to be of the form
\begin{equation}
\ket{\Psi} =  \frac{1}{\sqrt{2N}} \sum_{j=1}^{N} \left(e^{ik_nj} + e^{-ik_n(j-1)}\right) \ket{j},
\label{eq:CoeffLowFilling}
\end{equation}
where the quasi-momenta $k_n$ are quantised as
\begin{equation}
k_n = \frac{\pi n}{N} \: , \: n = 0,1,\dots,N-1 \: .
\label{eq:SpectraLowFilling}
\end{equation}
The spectrum is given by
\begin{equation}
K_n = U - 2t \cos(k_n).
\label{eq:EnergiesLowFilling}
\end{equation}
Notice, in our case, the quasi-momentum $k=0$ is allowed, resulting in the lowest $K_n$ state having a truly homogeneous density in the chain. Similar forms of the spectrum and quasi-momenta are found by solving the strongly-interacting one-dimensional Fermi gas with hard wall boundaries by Bethe ansatz \cite{Oelkers2006}.

\begin{figure}[t]
\begin{center}
\includegraphics[width=0.48\textwidth]{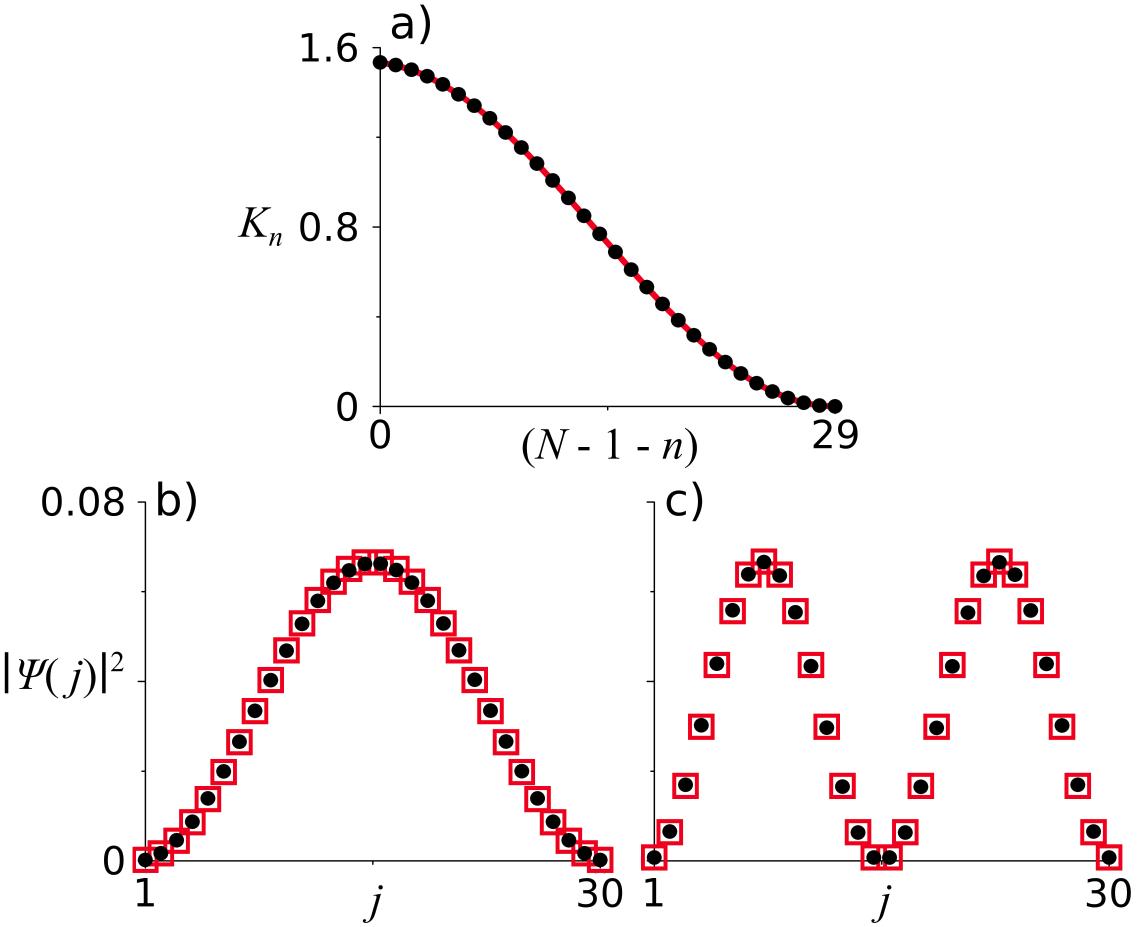}
\end{center}
\caption{Comparison of exact diagonalization of CONAN coefficients shown by circles (black) and the analytical model, Eqs.~(\ref{eq:WellWavefunctions}) and (\ref{eq:WellEnergy}), given by a solid line (red) in the spectrum and squares (red) for the probability densities for $\nu=1/2$, $N=30$. a) The energy spectrum, ordered so that the groundstate of the system (the highest $K_n$ state) corresponds to the index zero, b) groundstate ($N-1-n = 0$) probability density, c) first excited state ($N-1-n = 1$) probability density.}
\label{fig:HalfFilling}
\end{figure}

For the range $1/2 \leq \nu \leq 1$, we calculate the spin chain coefficients for each $L_s$ using the open source code CONAN. Using these coefficients we then solve the stationary Schr\"{o}dinger equation for Hamiltonian~(\ref{eq:KHam}) by exact diagonalization and compare the eigenfunctions and spectrum to Eqs.~(\ref{eq:CoeffLowFilling}) and~(\ref{eq:EnergiesLowFilling}) respectively. We find perfect agreement between the numerical and analytical results for $\nu < 1$, as shown for $\nu = 1/2$ in Fig.~\ref{fig:HalfFilling}. However, as the filling approaches $\nu = 1$ we, of course, observe small deviations due to finite size effects.

\subsection{High filling regime}
\label{sec:HighFill}

As we increase the filling beyond $\nu=1$, the physics of the single spin-down fermion change qualitatively in a dramatic way. As discussed in Sec.~\ref{sec:SpinCoefficients}, this is the result of the $N$ lowest energy single particle states filling the lowest band and partially occupying the second band for $\nu>1$. The form of the spin chain coefficients was discussed in Sec.~\ref{sec:SpinCoefficients} and is shown in Figs.~\ref{fig:CoefficientsExample}b and~\ref{fig:CoefficientsExample}c. The spin chain coefficients for $1< \nu < 2$ take the form of $p$ inverted finite wells, with the well height being small in some cases. As a result, both the analogous tight-binding model hopping and on-site potential acquire an oscillatory, inhomogeneous behaviour consisting of $p$ wells. Using the mappings of Eqs.~(\ref{eq:MappingTunnelingTightBinding}) and~(\ref{eq:MappingPotnentialTightBinding}), the potential takes the form of an inverted finite well, and the hopping a finite well of smaller magnitude.

We develop an approximate theory for the case of $p = 1$, which can be extended to multiple wells, assuming no coupling between each well. We expand the Hamiltonian parameters around the centre of the well and account for the hopping via an effective mass that is dependent on the hopping strength. Ref.~\cite{Valiente2008} considers a similar derivation for the case of a 1D Bose-Hubbard model. For a full derivation see Appendix~\ref{App:HighFilling}. Following these approximations, we obtain the eigenstates to be of the form~\cite{Valiente2008}
\begin{equation}
\begin{aligned}
\ket{\Psi} \approx & \: \sum_{z=-\frac{N+1}{2}}^{\frac{N+1}{2}} \frac{\mathcal{N}}{\sqrt{2^s s!}} \Bigg[ H_s\left(\sqrt[4]{\frac{\mathcal{U}}{\mathcal{T}}} z\right) \\ & \; \times e^{-\sqrt[4]{\frac{\mathcal{U}}{\mathcal{T}}}\frac{z^2}{2}} (-1)^z \bigket{z+\frac{N+1}{2}} \Bigg]
\end{aligned}
\label{eq:WellWavefunctions}
\end{equation}
with $\mathcal{U}$ and $\mathcal{T}$ being characterising variables for the potential and hopping strengths respectively, $\mathcal{N}$ the normalisation constant, $H_s$ the $s$th Hermite polynomial, $z = j - (N+1)/2$ and $s = 0,1,\dots,(N-1)$. The characterising parameters of $\mathcal{U}$ and $\mathcal{T}$ are defined in Appendix~\ref{App:HighFilling}. The solutions are that of the harmonic oscillator with a $\pi$-phase term. The approximation to the energy spectrum is given by \cite{Valiente2008}
\begin{equation}
K_s \approx E_{\mathrm{off}} - 2 \sqrt{\mathcal{T U}}\left(s + \frac{1}{2}\right)
\label{eq:WellEnergy},
\end{equation}
with $E_{\mathrm{off}}$ being a constant offset to the energy that is calculated during the reduction of the hopping in the effective mass approximation.

\begin{figure}[t]
\begin{center}
\includegraphics[width=0.49\textwidth]{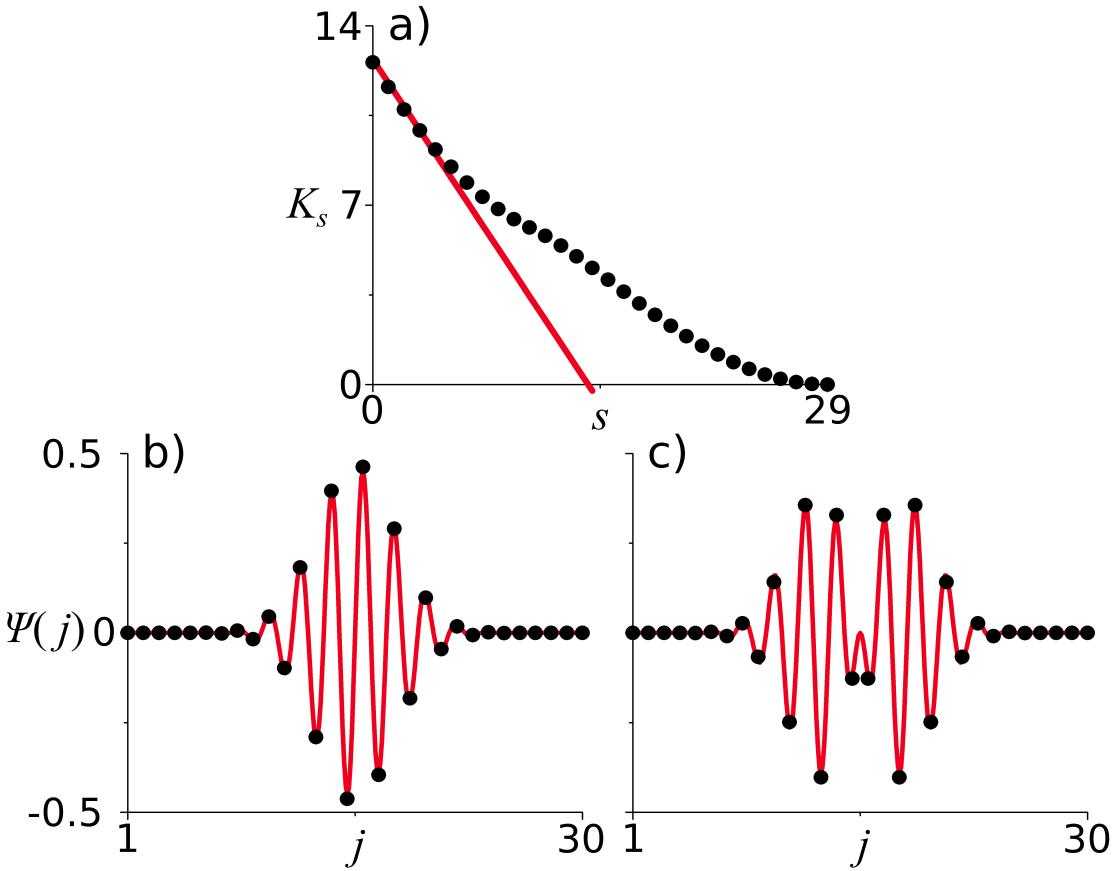}
\end{center}
\caption{Comparison of exact diagonalization of CONAN coefficients shown by circles (black) and the analytical model, Eqs.~(\ref{eq:WellWavefunctions}) and (\ref{eq:WellEnergy}), given by a solid line (red) for $\nu=1+1/N=1.033$, $N=30$. a) The energy spectrum, b) groundstate ($s = 0$) eigenfunction, c) first excited state ($s = 1$) eigenfunction.}
\label{fig:WellApproximation}
\end{figure}

In Fig.~\ref{fig:WellApproximation}, we compare the analytical eigenfunctions and eigenvalues (Eqs.~(\ref{eq:WellWavefunctions}) and~(\ref{eq:WellEnergy}) respectively) to the numerical exact diagonalization of the spin chain Hamiltonian with the coefficients calculated by CONAN for $L_s = 29$, $N = 30$. There is an excellent agreement at low energies (high $K_s$). As expected, due to the finite depth of the well, we recover plane wave solutions like that for $\nu \leq 1$ for high energies.

\subsection{Double filling}
\label{sec:Double}

\begin{figure}[t]
\begin{center}
\includegraphics[width=0.45\textwidth]{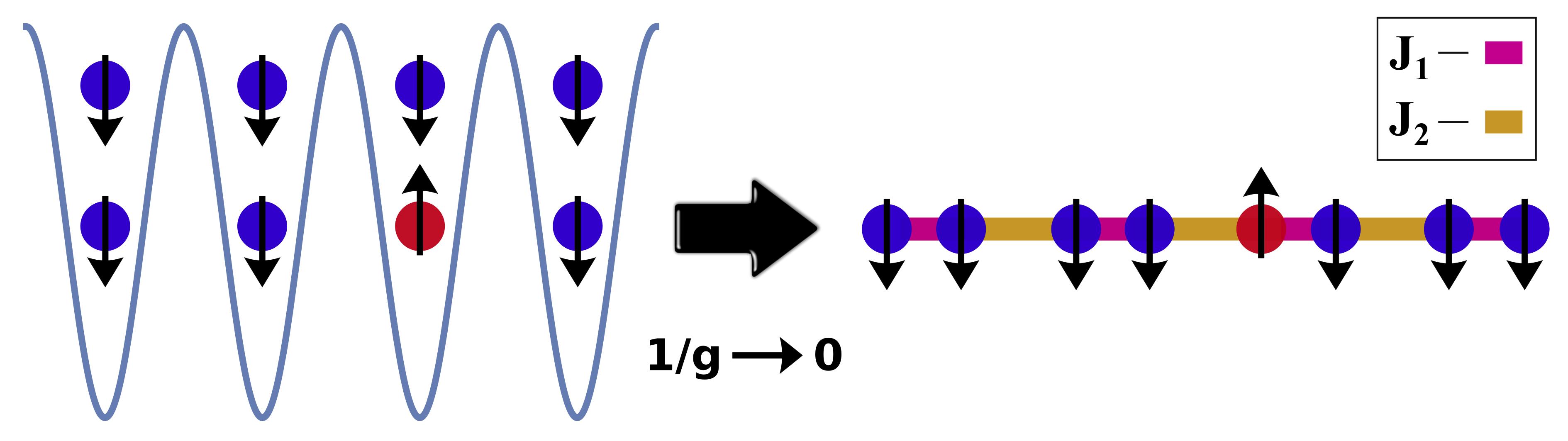}
\end{center}
\caption{Illustration of the mapping of the system to a spin chain when $g \rightarrow \infty$ at double filling ($\nu=2$) with coupling constants $J_1$ and $J_2$.}
\label{fig:DoubleFillingModel}
\end{figure}

At $\nu = 2$ every well of the optical lattice is doubly occupied, and there are $N/2$ states occupied in each of the first two bands of the single particle spectrum. This results in a distortion of the spin chain similar to that of the Peierls transition \cite{Peierls1955}, see Fig.~\ref{fig:DoubleFillingModel}. From this, it would be expected that the spin chain coefficients are staggered between two values of intra- and inter-well couplings, as illustrated in Fig.~\ref{fig:DoubleFillingModel}, with the coefficients form given in Fig.~\ref{fig:CoefficientsExample}d and Fig.~\ref{fig:CoefficientsExampleN20}d. The spin chain coefficients between particles in a single well of the optical lattice are naturally larger than that between particles in separate wells. The staggering in the spin chain coefficients is a result of two atoms sitting in single wells of the optical lattice potential, which will occur for any lattice length as long as there is double filling.

For this special point, the single particle tight-binding model on-site potential for $j \neq 1, N$ is essentially constant and the hopping is staggered between two values. This form of the single particle Hamiltonian is similar to that of the Su-Schriefer-Heeger (SSH) model for polyacetylene \cite{Su1979,Su1980}. The SSH model has a unit cell comprising of two sites (A and B), and its Hamiltonian has the form \cite{Heeger1988,Asboth2016}
\begin{equation}
\begin{aligned}
\hat{H}_{\mathrm{SSH}} = & \sum_{u=1}^{N/2} (t + \delta t) \left[ \hat{b}_{\mathrm{A},u}^{\dagger} \hat{b}_{\mathrm{B},u}^{\:} + \mathrm{h.c. } \right] \\ & + \sum_{u=1}^{N/2-1} (t - \delta t) \left[ \hat{b}_{\mathrm{A},u+1}^{\dagger} \hat{b}_{\mathrm{B},u}^{\:}  + \mathrm{h.c. }  \right],
\end{aligned}
\label{eq:SSHHamiltonian}
\end{equation} 
where $\hat{b}_{A(B),v}^{\dagger}$ and $\hat{b}_{\mathrm{A(B)},v}$ act on the A(B) sub-lattice site of the unit cell $u$. With hopping coefficients of $(t+\delta t)$ within each unit cell and $(t - \delta t)$ between adjacent unit cells. To compare to the numerics we find the average $t$ and $\delta t$ across the whole spin chain. This model has two bands, which we label $\pm$. The values obtained for $t$ and $\delta t$ are dependent on the lattice strength. To increase the ratio $\delta t / t$, an increase in the lattice strength is required. This can be motivated by the diagram of Fig.~\ref{fig:DoubleFillingModel}, where it can be visualised that increasing the lattice strength would in effect decrease the coupling between atoms in different lattice wells, giving a corresponding increase in the ratio $\delta t / t$.

By observing the spin chain coefficeint data in Fig.~\ref{fig:CoefficientsExample}d and Fig.~\ref{fig:CoefficientsExampleN20}d, we can treat the single particle on-site potential as a constant with a strength of $2t$, that is, we assume that the physics is dominated by the staggered tunnelling over the inhomogeneity of the potential at the edges. The analogous single particle Hamiltonian at $\nu = 2$ is then of the approximate form,
\begin{equation}
\hat{K}^{(1)} = \sum_{u=1}^{N/2} 2 t (\hat{b}^{\dagger}_{A,u} \hat{b}_{A,u} + \hat{b}^{\dagger}_{B,u} \hat{b}_{B,u}) - \hat{H}_{\mathrm{SSH}}
\label{eq:SingleParticleSSH}.
\end{equation}

In Appendix~\ref{App:DoubleFilling} we solve Hamiltonian~(\ref{eq:SingleParticleSSH}) for the eigenstates and spectrum. Quoting the results of this derivation, the eigenstates are given by
\begin{equation}
\ket{\Psi} = \mathcal{N} \sum_{j=1}^N \left[ \phi_k(j) e^{i k j} - \phi_{-k}(j) e^{-i k j} \right] \ket{j},
\label{eq:wavefunctionDoubleFilling}
\end{equation}
where $\mathcal{N}$ is a normalisation factor, and $\phi$ is the Bloch function of the SSH Hamiltonian~(\ref{eq:SSHHamiltonian}),
\begin{equation}
\phi_k(x) = \begin{pmatrix}
\pm e^{i \kappa(k)} \\ 1
\end{pmatrix},
\label{eq:SSHSolution}
\end{equation}
where the $\pm$ refers to the two bands of the model. The Bloch functions are periodic, $\phi_k(x + 2) = \phi_k (x)$, and the phase $\kappa$ in Eq.~(\ref{eq:SSHSolution}) is given by
\begin{equation}
\kappa (k) = \arctan \left( \frac{\delta t}{t} \tan \left( k \right) \right).
\label{eq:SSHPhase}
\end{equation}
The allowed quasi-momenta are obtained from the open boundary conditions and are given by
\begin{equation}
k = \frac{1}{N+1} \left[ n \pi - \kappa(k) \right] \: , \: n = 1,2,\dots,\frac{N}{2}
\label{eq:MomentaDoubleFilling}.
\end{equation}
The quasi-energy for Hamiltonian~(\ref{eq:SingleParticleSSH}) is
\begin{equation}
K_n = 2t \mp 2 \sqrt{t^2 \cos^2 (k) + \delta t^2 \sin^2 (k)}.
\label{eq:EnergiesDoubleFilling}
\end{equation}

\begin{figure}[t]
\begin{center}
\includegraphics[width=0.49\textwidth]{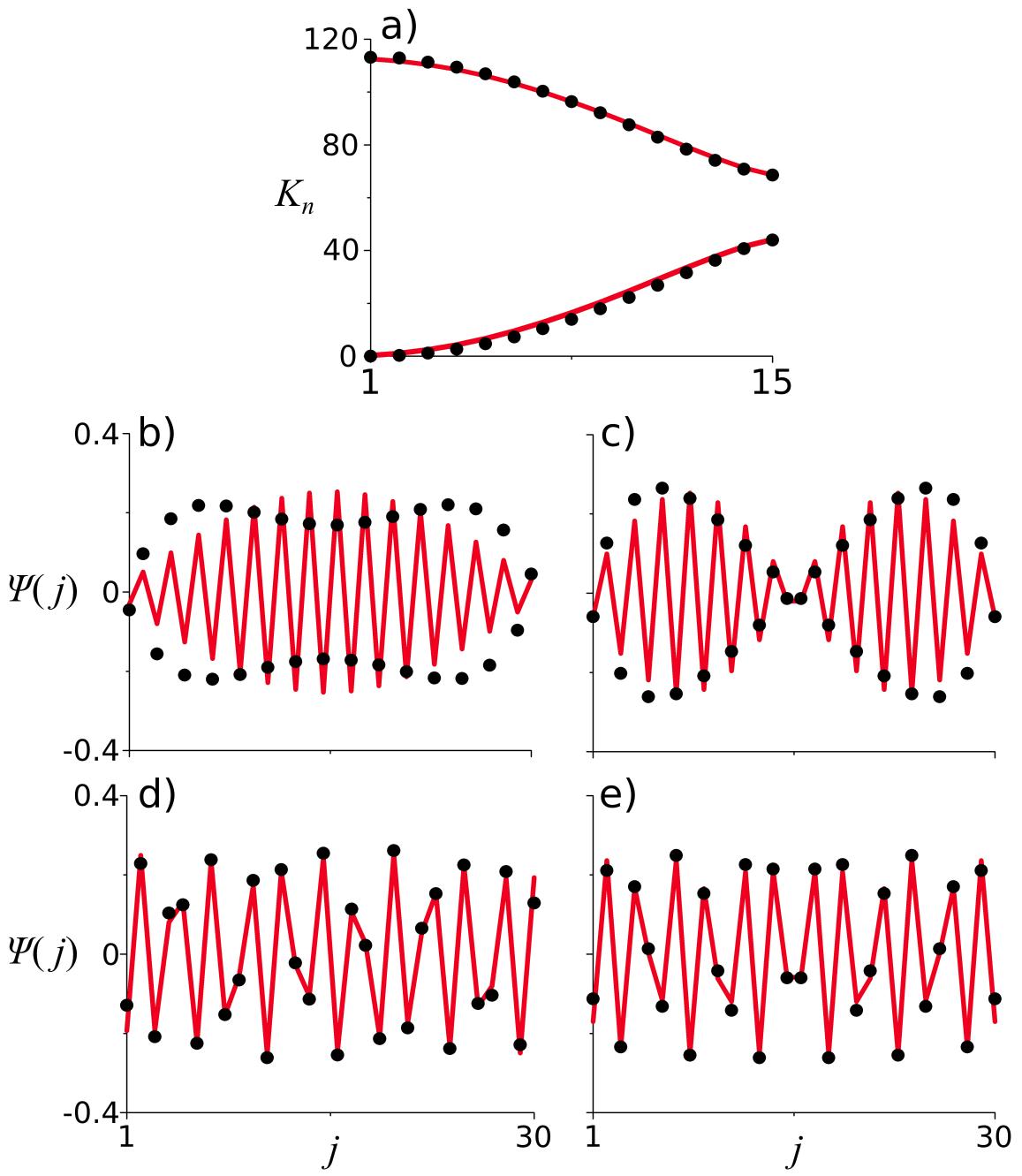}
\end{center}
\caption{Comparison of exact diagonalization of CONAN coefficients shown by circles (black) and the analytical model, Eqs.~(\ref{eq:wavefunctionDoubleFilling}) and (\ref{eq:EnergiesDoubleFilling}), given by a solid line (red) for $\nu = 2$, $N = 30$. a) The energy spectrum in the reduced zone scheme, b--f) eigenfunctions for the b) groundstate of system (highest $K_n$), c) first excited state, e) fifth excited state, f) sixth excited state.}
\label{fig:DoubleSSHComparison}
\end{figure}

In Fig.~\ref{fig:DoubleSSHComparison}, we compare the analytical eigenfunctions and spectrum (Eqs.~(\ref{eq:wavefunctionDoubleFilling}) and (\ref{eq:EnergiesDoubleFilling}) respectively) to the results from solving the stationary Schr\"{o}dinger equation for Hamiltonian~(\ref{eq:KHam}) by exact diagonalization using the spin chain coefficients calculated by CONAN for $\nu = 2$. We find excellent agreement for the spectrum of the model. For all but the groundstate, we observe good agreement between the analytical and numerical eigenfunctions, see Fig.~\ref{fig:DoubleSSHComparison}c, d and e. For the groundstate of the system, the highest $K_n$ state, we observe a substantial discrepancy between the analytical eigenfunctions and the exact diagonalization due to finite size effects, seen in Fig.~\ref{fig:DoubleSSHComparison}b. The finite size of the system results in a small well-like perturbation to the intra-well coupling $(t + \delta t)$ of order $10^{-6} \mathrm{max}(J_j) z^4$, again $z = j - (N+1)/2$. This perturbation can be seen in the form of the spin chain coefficients shown in Fig.~\ref{fig:CoefficientsExample}d, with the decrease of order $0.05 \: \mathrm{max}(J_j)$ in the centre of the chain for the larger of the staggered coefficients. Unsurprisingly, this deviation results in the significant modification of the low energy, high $K_n$, states, most specifically to the groundstate.

\section{Magnon-magnon Interactions in Multi-impurity Systems}
\label{sec:Interactions}

\subsection{Two Magnons}

\begin{figure}[t]
\begin{center}
\includegraphics[width=0.48\textwidth]{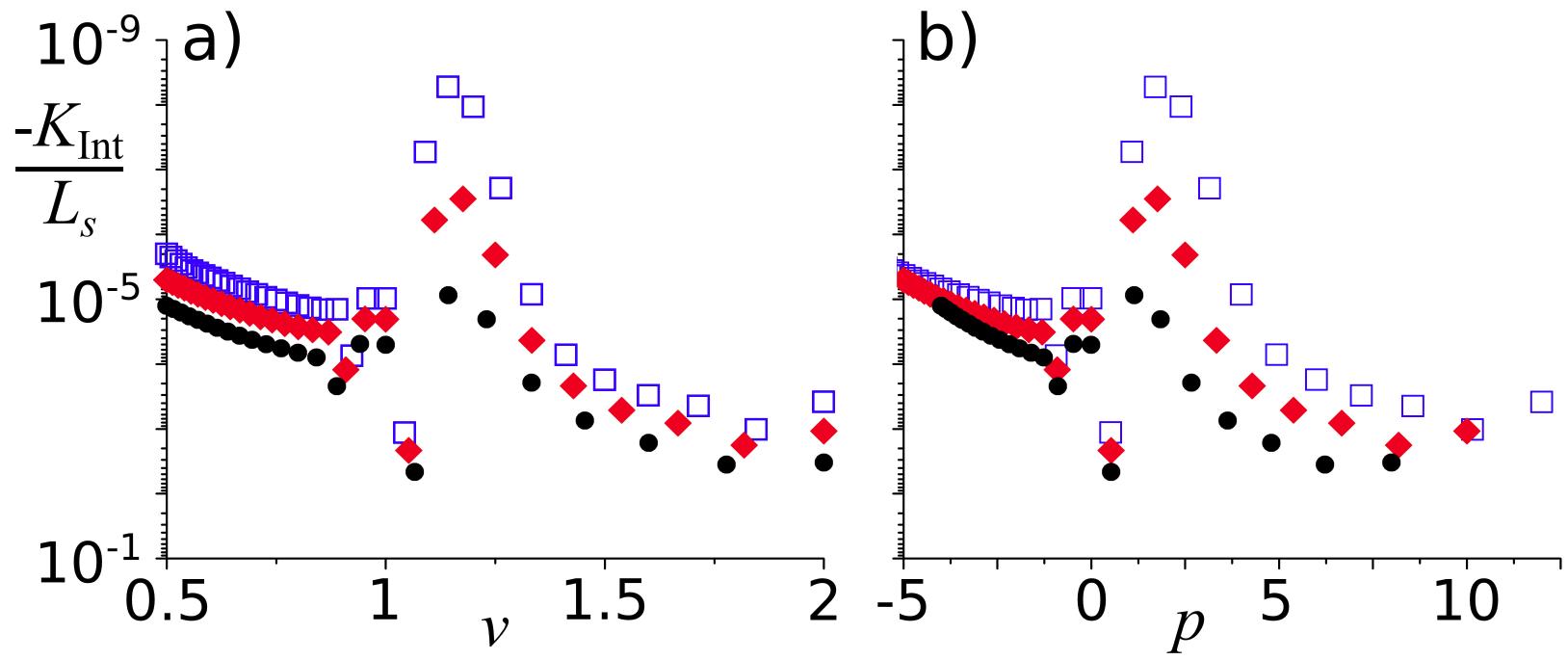}
\end{center}
\caption{Magnon--magnon energy shift, Eq.~(\ref{eq:KInt}), for two spin-down fermions on a log--scale for $N=16$ circles (black), $N=20$ diamonds (red), and $N=24$ squares (blue outline). a) as a function of filling $\nu$ and b) a function of the number of particles over unit filling $p$.}
\label{fig:TwoPolaron}
\end{figure}

We now move on to discuss the effects of interaction between magnons within the spin chain. We calculate the ground state energy for two spin-down fermions, which we denote $K_0^{\mathrm{I}}$, by exact diagonalization across the three regimes previously discussed and quantify the interactions between the impurities by the magnon-magnon energy shift, 
\begin{align}
K_{\mathrm{Int}} = K_0^{\mathrm{I}} - K_0^{\mathrm{NI}} ,
\label{eq:KInt}
\end{align}
with $K_0^{\mathrm{NI}}$ being the non-interacting ground state for two hard-core bosons (free magnons), i.e. the sum of the ground and first excited single magnon energies.

To support the discussion we will refer to the calculated eigenfunctions for the two magnons. We write the magnon-magnon wavefunction in the basis of states $\ket{j_1,j_2}$, with $j_i$ denoting the position of the $i\mathrm{th}$ spin, i.e.
\begin{align}
\Psi = \sum_{j_1 < j_2} \psi \left( j_1, j_2 \right) \ket{j_1,j_2}.
\end{align}
In plotting the eigenfunctions, we plot the coefficients $\psi \left( j_1, j_2 \right)$, with a mirror image across the line $j_1 = j_2$.

In Fig.~\ref{fig:TwoPolaron}a, we show $K_{\mathrm{Int}}$ as a function of the filling $\nu$ for $N = 16,20,24$, with $\nu$ scaled through by varying the number of lattice sites $L_s$. To see the behaviour of the interactions clearly we also plot, in Fig.~\ref{fig:TwoPolaron}b, $K_{\mathrm{Int}}$ as a function of the number of particles off of unit filling $p$, i.e. $\nu = N/ \left(N - p\right)$. The magnon-magnon interaction shift is found to be attractive in all cases.

\begin{figure}[t]
\begin{center}
\includegraphics[width=0.49\textwidth]{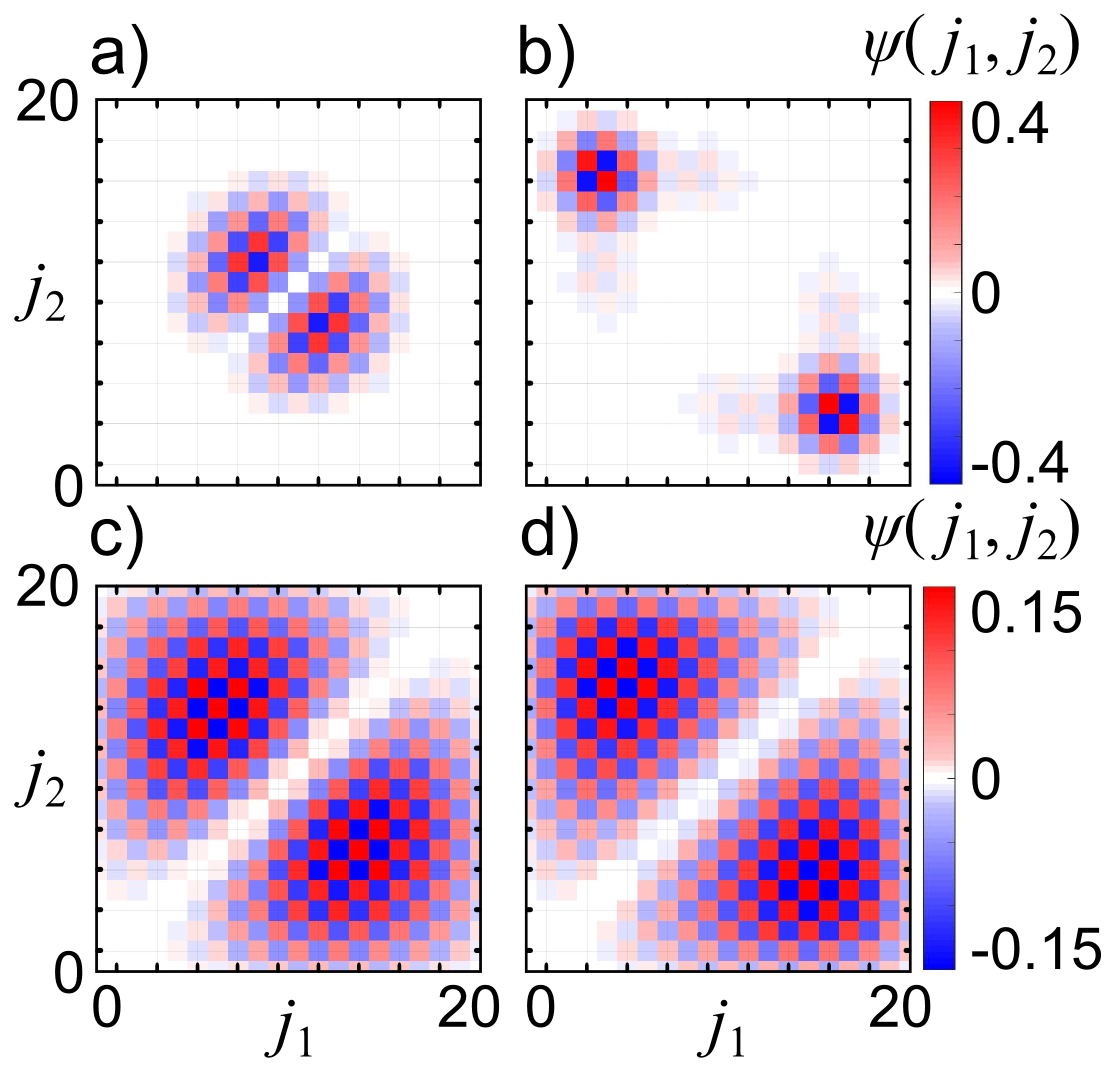}
\end{center}
\caption{Eigenfunctions for the case of two impurities (magnon-magnon) $\psi \left(j_1,j_2\right)$, which is the coefficient to the basis state $\ket{j_1,j_2}$. All plots are for $N = 20$ and a filling of a) $\nu = 1.053$ ($p = 1$), b) $\nu = 1.176$ ($p = 3$), c) $\nu = 1.818$ ($p=9$) and d) $\nu = 2$ ($p=10$).}
\label{fig:Eigenfunctions}
\end{figure}

We observe a clear transition at $\nu = 1$($p = 0$). This is to be expected from our investigation of the single impurity system. This is a transition from a homogeneous dilute regime, to a regime with a localising `well-like' potential. We observe an increase in the interaction for the single well case, $p=1$, as the two magnons are localised in close proximity of each other. This is clear in the eigenfunction Fig.~\ref{fig:Eigenfunctions}a, with the state heavily localised to a region where the two magnons are close together. For $p=2$ there is a substantial decrease in the interaction energy shift, due to the two magnons being spatially separated by the double well form of the spin chain coefficients. This is observed for the case of $p=3$ in the eigenfunction of Fig.~\ref{fig:Eigenfunctions}b, with the state heavily localized to a region where the two magnons are well separated. As $p$ is increased the approximation of multiple wells with no coupling between them breaks down, resulting in an increase in the interaction strength as observed with the magnons delocalised across the system as seen in Fig.~\ref{fig:Eigenfunctions}c.

We observe a pronounced decrease of the interaction strength for all $N$ at double filling, $\nu = 2$. As already discussed in Sec.~\ref{sec:Double}, at this point, excluding finite size effects, we have an effective SSH model for the system. This results in a reduced interaction of the two magnons. This can be seen in the eigenfunctions, with a shift in the distribution of the coefficients towards the magnons being further apart. That is, considering the lower triangle half of Figs.~\ref{fig:Eigenfunctions}c and d, there is a small shift in the coefficients of the basis towards values of the magnons being further apart, i.e. towards the point of $\left( j_1,j_2 \right) = \left( 20,0 \right)$, resulting in the decrease in interaction strength.

\subsection{Spin Balanced Case}

Using the density-matrix-renormalization-group (DMRG), we obtain the ground state energies, $K_0^{\mathrm{I}}$, for the case of balanced spins $N_{\downarrow} = N_{\uparrow} = N/2$ in the spin chain. We again calculate $K_{\mathrm{Int}}$, given by Eq.~(\ref{eq:KInt}), which is now an $N/2$ magnon energy shift. The non-interacting groundstate $K_0^{\mathrm{NI}}$ is in this case the sum of the first $N/2$ single magnons energies, i.e. the highest $N/2$ states in $K_n$.

\begin{figure}[t]
\begin{center}
\includegraphics[width=0.48\textwidth]{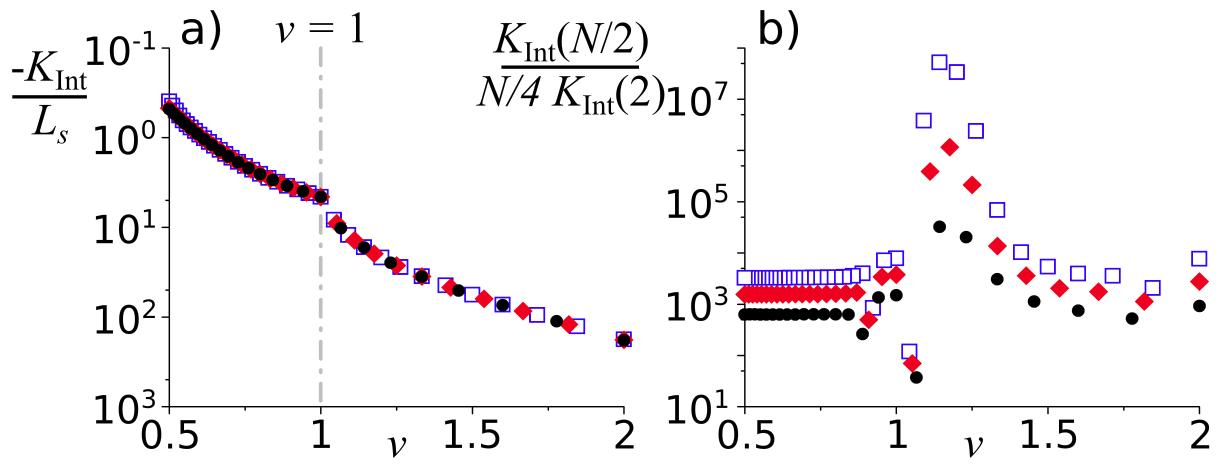}
\end{center}
\caption{a) Energy shift due to interactions on a log--scale with $N=16$ circles (black), $N=20$ diamonds (red), and $N=24$ squares (blue outline) for the spin balanced case $N_{\downarrow} = N_{\uparrow} = N/2$. b) $K_{\mathrm{Int}}({N/2})/(N/4 \: \: K_{\mathrm{Int}}(2))$, the ratio of the balanced interaction strength to the two spin-down fermion case for the same $N$ considered in a).}
\label{fig:InteractionBalanced}
\end{figure}

In Fig.~\ref{fig:InteractionBalanced}a we observe a collapse of the transition observed for two magnons for all $N$. The collapse is due to the fixed proportion of magnons in the system as the system size is altered. The transition is now, in the fermionic language for magnons, a standard metal-insulator-metal transition. The discontinuity of the energy shift at the transition has a magnitude of $\Delta K_{\mathrm{Int}}/L_s \approx 5$.

For $\nu < 1$, the interaction energy shifts of the $N/2$ magnons are well described by the two-magnon energy shifts, as can be seen by the approximately constant ratio of the balanced spin and two magnon case for fixed $N$, $K_{\mathrm{Int}}({N/2})/(N/4 \: \: K_{\mathrm{Int}}(2))$ in Fig.~\ref{fig:InteractionBalanced}b. This means that for $\nu < 1$ the interaction is a result of mainly two-body processes. As the filling approaches unity the nature of the interactions goes through a transition, due to the behaviour of the two magnons system discussed in the previous section. Overall for $\nu > 1$ the nature of the interactions is not well described by only two-body processes, due to the formation of `well-like' structure in the Hamiltonian as discussed previously.

\section{Conclusions}

In this work, we have considered the limit of both strong repulsive interactions and one dimension of the Fermi polaron problem in an optical lattice. In this limit polaronic-behaviour is not expected, instead, we find three distinct regimes for the impurity, or magnon, over a reasonable range of fillings. For low filling, $\nu \leq 1$ the magnon reproduces a single particle tight-binding model. With high filling, $\nu > 1$, we observe a localization of the low energy (high $K_n$) single magnon eigenfunctions,  due to a `well-like' form to the Hamiltonian, which is a direct result of the occupation of the second band of the lattice and the moderate strength of the optical lattice. The eigenfunctions at low energy in this regime reproduce the harmonic well eigenfunctions with a $\pi$-phase term. The final regime occurs at the point of double filling, $\nu = 2$, where the spin chain coefficients are of a staggered form. The Hamiltonian, when written as an analogous tight binding model, takes the form of the well-known SSH model. For all but the lowest state of the spectrum, the single magnon eigenfunctions reproduce that of the SSH model. However, at low energies the eigenfunctions have adverse finite size effects that result in discrepancies, most significantly to the groundstate of the system (highest $K_n$ state).

In the final section of this work, we consider the nature of the interactions of multiple spin-down fermions by considering two magnons and the important spin balanced case. We observe a rich transition reflecting the three regimes present. There is a clear metal-insulator-metal transition as the filling is increased, which corresponds to the three regimes present for the single magnon.

In summary, we have shown that a single spin impurity in a spin-1/2 Fermi gas within an optical lattice potential in the limit of one dimension and strong repulsive interactions can have a rich set of behaviour dependent on the filling of the lattice. The system is found to replicate the quantum models of homogeneous systems, finite wells and the Su-Schriefer-Heeger model.

\acknowledgements{C.W.D. acknowledges support from EPSRC CM-CDT Grant No. EP/L015110/1. P.\"O. and M.V. acknowledge support from EPSRC EP/M024636/1. F. F. B. and N. T. Z. acknowledge support by the Danish Council for Independent Research DFF Natural Sciences and the DFF Sapere Aude program.}

\appendix

\section{Tight-binding Model Analogy}
\label{app:TightBinding}

The spin chain Hamiltonian, Eq.~(\ref{eq:KHam}), for a single spin-down fermion can be written as a matrix in the basis of $\ket{j}$ ($j$ denotes the position of the spin-down fermion). For example with $N=4$ we have
\begin{equation}
\hat{K} = \begin{pmatrix}
J_{1} & -J_{1} & 0 & 0 \\
-J_{1} & J_{1}+J_{2} & -J_{2} & 0  \\
0 & -J_{2} & J_{2}+J_{3} & -J_{3}\\
0 & 0 & -J_{3} & J_{3} \\
\end{pmatrix},
\label{eq:PolaronHamiltonian}
\end{equation}
with $J_j$ denoting the spin chain coefficient between the $j$ and $j+1$ site of the spin chain. Note that due to the hard-wall boundaries the edge sites of the chain are only coupled to one other site. This results in an inhomogeneity at the edges of the spin chain which has interesting effects on the system.

We can trivially map the Hamiltonian for a single spin-down fermion to that of a single particle tight-binding model, as discussed in the main text with the analogous Hamiltonian given in Eq.~(\ref{eq:ParticleHam}). Looking at the example of Eq.~(\ref{eq:PolaronHamiltonian}), we simply have a set of diagonal terms denoting an on-site potential, $U$, and off-diagonal terms denoting hopping strength between nearest-neighbours, $t$. This results in the mapping between spin chain coefficient and the tight-binding model given in Eqs.~(\ref{eq:MappingTunnelingTightBinding}) and~(\ref{eq:MappingPotnentialTightBinding}) in the main text.

\section{Form of the Effective Spin Chain}
\label{App:Coefficients}

In Sec.~\ref{sec:SpinCoefficients}, we discuss the three regimes of the single-particle system as a function of filling in the optical lattice potential. These three regimes, when the strong interaction limit is taken, result in stark differences in behaviour for the spin chain coefficients of Hamiltonian~(\ref{eq:KHam}). In this appendix we give a brief discussion of the form of the spin chain coefficients across the full region considered, $1/2 \leq \nu \leq 2$, as computed by the open source code CONAN \cite{Loft2016a} for $N = 20$.

\begin{figure}[t]
\begin{center}
\includegraphics[width=0.45\textwidth]{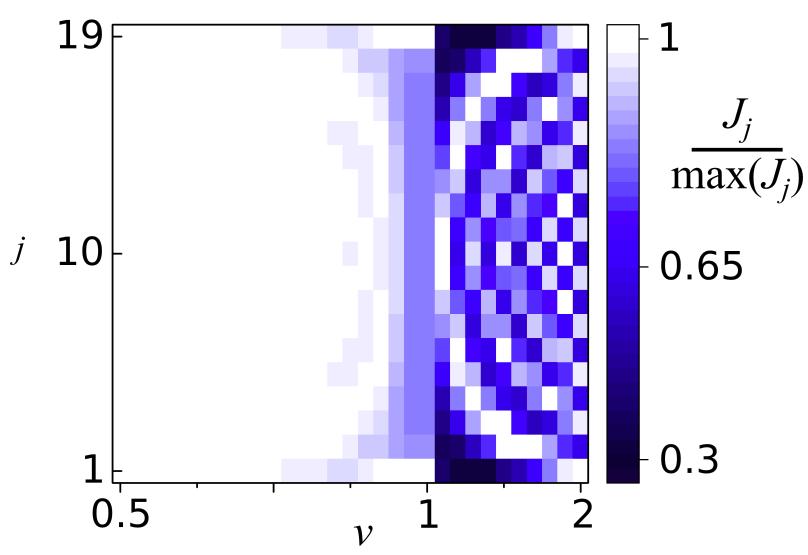}
\end{center}
\caption{Normalised spin chain coefficients, $J_j$, for $N=20$ over the range of fillings $\nu$ considered.}
\label{fig:CouplingConstants}
\end{figure}

As expected we observe three distinct regimes across the range of fillings considered, see Fig.~\ref{fig:CouplingConstants}. For $\nu \leq 1$, there is a homogeneous regime, with finite size effects when approaching $\nu = 1$. The transition to the inverted `well-like' forms of $1 < \nu < 2$ is sharp. As the filling is increased above unity,  i.e. $\nu = N/ \left(N - p\right)$, for each $p$ there can be seen to be $p$ `wells' in the form of the coefficients. Between the second and third regime, there is an extended transition from about $p \sim 5$ until the special point of $\nu = 2$. Finally, at double filling, $\nu = 2$, the coefficients take on a staggering form between intra- and inter-well couplings between particles.

\section{Low Filling Derivation}
\label{App:LowFilling}

In this regime, we have a clear consistent form to the spin chain coefficients, see Fig.~\ref{fig:CoefficientsExample}a. Mapping the coefficients to the analogous single particle tight-binding Hamiltonian, see Eqs.~(\ref{eq:MappingTunnelingTightBinding}) and~(\ref{eq:MappingPotnentialTightBinding}), there is a constant negative hopping of value $-t$ and an on-site potential that is constant for $j \neq 1, N$ at a value of $U = 2t$, with $U,t > 0$. At the sites $j=1,N$ we have the on-site potential equal to $t$.

The analogous tight-binding Hamiltonian in this regime is given by
\begin{equation}
\begin{aligned}
\hat{K} = & -t \sum_{j=1}^{N}  \left( \hat{b}^{\dagger}_j \hat{b}_{j+1} + \hat{b}^{\dagger}_{j+1} \hat{b}_{j} \right) + 2t \sum_{j=2}^{N-1} \hat{b}^{\dagger}_j \hat{b}_j \\ & + t \left( \hat{b}^{\dagger}_1 \hat{b}_1 + \hat{b}^{\dagger}_N \hat{b}_N \right),
\end{aligned}
\label{eq:LowFillingHamiltonian}
\end{equation}
with $\hat{b}_j(\hat{b}_j^{\dagger})$ denoting the annihilation(creation) operator at site $j$. Excluding the sites $j = 1,N$, we have a simple tight-binding Hamiltonian and we would expect plane wave solutions at site $j$, with $j \neq 1,N$, of the form
\begin{equation}
C_j = e^{ikj}+B_j e^{-ikj},
\label{eq:PlaneWave}
\end{equation}
where k is the momentum and B a coefficient to be determined. Writing the stationary Schr\"{o}dinger equation of Hamiltonian~(\ref{eq:LowFillingHamiltonian}), for $j \neq 1,N$ in a discretized form we obtain
\begin{equation}
-2t \left( C_{j+1} + C_{j-1} \right) + t C_j = K_n C_j .
\label{eq:SmallFillingHam}
\end{equation}
The spectrum of this homogeneous system is
\begin{equation}
K_n = 2t - 2t\cos\left(k\right).
\end{equation}

We now derive the exact form of $C_j$, and the quasi-momentum $k$. We can solve for $B_1$ and $C_1$ by considering $j = 1$ and $j=2$ for the discretized Hamiltonian Eq.~(\ref{eq:SmallFillingHam}). Solving these results in
\begin{equation}
\begin{aligned}
B_1 &= e^{ik} \\
C_1 &= 1 + e^{ik}.
\end{aligned}
\end{equation}
We can similarly find $B_N$ and $C_N$ using $j=N$ and $j=N-1$, giving
\begin{equation}
\begin{aligned}
B_N &= e^{ik(2N+1)} \\
C_N &= e^{ikN}\left(1 + e^{ik}\right).
\end{aligned}
\end{equation}
Setting $B_1 = B_N$, we get the quasi-momenta to be quantised as
\begin{equation}
k = \frac{\pi n}{N} \: , \: n \in \mathbb{Z}
\label{eqn:MomentumPeriodic}
\end{equation}
therefore $n=0,1,\dots,N-1$. Finally, from this derivation, we find a form of the general $jth$ coefficient and find the normalised eigenfunctions to be given by
\begin{equation}
\ket{\Psi_n} = \frac{1}{\sqrt{2N}} \sum_{j=1}^{N} \left( e^{i k_n j} + e^{-i k_n \left(j-1\right)} \right) \ket{j}.
\end{equation}

\section{High Filling Derivation}
\label{App:HighFilling}

For $\nu > 1$, the analogous tight-binding Hamiltonian has a finite `well-like' structure to both the potential and hopping, as discussed in Sec.~\ref{sec:HighFill}. In this appendix, we derive approximate eigenfunctions and spectrum for the case of one particle over unit filling, $\nu = 1 + 1/L_s$, corresponding to a single well in the tunnelling coefficients.

To make the derivation simpler, we centre the spin chain sites at the origin, that is $j \rightarrow j - (N+1)/2$, and we will label this coordinate as $z$. For the analogous tight-binding Hamiltonian we can approximate the potential and hopping as  
\begin{equation}
\begin{aligned}
U_z = & U^0 + U^1 \cos\left(\tau z\right) \\
t_z = & -t^0 - t^1 \cos\left[\tau \left(z + \frac{1}{2}\right)\right],
\end{aligned}
\label{eq:ApproxWells}
\end{equation}
with $\tau = 2 \pi / (N-3)$ the spatial frequency, and $U^0$, $U^1$, $t^0$ and $t^1$ are positive ($> 0$) constants, with each found by fitting Eqs.~(\ref{eq:ApproxWells}) to the analogous potential and hopping obtained from the mapping of the spin chain coefficients by Eqs.~(\ref{eq:MappingTunnelingTightBinding}) and~(\ref{eq:MappingPotnentialTightBinding}). Note, that the cosine term in Eqs.~(\ref{eq:ApproxWells}) allows for the finite size effect of decreased spin chain coefficients at the boundaries to be accounted for.

We solve the analogous tight-binding Hamiltonian (Eq.~(\ref{eq:ParticleHam}) for the potential and hopping of Eqs.~(\ref{eq:ApproxWells}). We can define the hopping as positive by considering a phase shift to the full eigenfunction of the effective form
\begin{equation}
\Psi_z = \psi_z (-1)^z,
\end{equation}
resulting in the single-particle Hamiltonian for $\psi_z$ being
\begin{equation}
\hat{K}^{(1)} = \sum_{z = - (N-1)/2}^{(N-1)/2} \left[ U_z \hat{b}^{\dagger}_{z} \hat{b}_z + t^{\prime}_z \left( \hat{b}^{\dagger}_{z} \hat{b}_{z+1} + \mathrm{h.c.}  \right) \right], 
\end{equation}
with $t^{\prime}_z = -t_z$.

Expanding to zeroth order around the centre of the system, $z=0$, we get a Hamiltonian of the form 
\begin{equation}
\begin{aligned}
\hat{K}^{(1)} \approx \sum_{z = - (N-1)/2}^{(N-1)/2} [ & (U^0 + U^1) \hat{b}^{\dagger}_{z} \hat{b}_z \\ & + (t^0 + t^1) \left( \hat{b}^{\dagger}_{z} \hat{b}_{z+1} + \mathrm{h.c.}  \right) ]. 
\end{aligned}
\end{equation}
This can be diagonalized by Fourier transform to obtain the eigenvalues and expanding for the low energy spectra around $k = 0$ we obtain
\begin{equation}
E_s \approx U^0 + U^1 + 2(t^0+t^1) - (t^0+t^1) k_s^2 \: ,
\label{eq:ZeroEigen}
\end{equation}
with $k_s$ being the momentum of state $s$. We can separate Eq.~(\ref{eq:ZeroEigen}) into two parts, with the first being a constant offset of
\begin{equation}
E_{\mathrm{off}} = U^0 + U^1 + 2(t^0+t^1),
\end{equation}
which will be accounted for in the final energy. The second term is related to the energy of a particle with an effective mass $m^{\star}$. Equating the last term in Eq.~(\ref{eq:ZeroEigen}) to $\hbar^2 k_s^2 / 2 m^{\star}$ gives an effective mass of
\begin{equation}
m^{\star} = - \frac{\hbar^2}{2(t^0 + t^1)}.
\end{equation}
Note, $t^0, t^1 > 0$, therefore, the effective mass is negative, as would be expected from the form of the on-site potential.

We can now write the Hamiltonian as an effective model of only a potential, with the hopping properties accounted for by the effective mass \cite{Valiente2008}. Taking the expansion around $z = 0$ of the potential to first order we obtain
\begin{equation}
U_z \approx U^0 + U^1 - \frac{1}{2} U^1 z^2 .
\label{eq:PotentialExpansion}
\end{equation}
Taking $x$ to be the continuous counterpart of $z$, we approximate the Hamiltonian to be
\begin{equation}
-\frac{\hbar^2}{2 m^{\prime}} \frac{d^2 \psi (x)}{dx^2} + \frac{1}{2} U^1 \tau^2 x^2 \psi(x) = - \epsilon \psi(x),
\label{eq:ContinuosHam}
\end{equation}
where we have defined the mass as negative $m^{\prime} = - m^{\star}$, and $\epsilon = E - E_{\mathrm{off}}$, with $E$ being the eigenvalues of Eq.~(\ref{eq:ContinuosHam}) without the offset. The form of Eq.~(\ref{eq:ContinuosHam}) is that of the well known harmonic oscillator, with an analogous frequency of
\begin{equation}
\omega^2 = \frac{U^1 \tau^2}{m^{\prime}}.
\end{equation}

From this we can derive two characteristic quantities of the harmonic well problem to be \cite{Valiente2008}
\begin{align}
\hbar w = & 2 \sqrt{\frac{1}{2} U^1 \alpha^2\left(t^0+t^1\right)} = 2 \sqrt{\mathcal{U} \mathcal{T}} \label{eq:Constants1}\\
\frac{m^{\prime} w}{\hbar} = & \sqrt{\frac{\frac{1}{2} U^1 \alpha^2}{\left(t^0+t^1\right)}} = \sqrt{\frac{\mathcal{U}}{\mathcal{T}}}.
\label{eq:Constants2}
\end{align}
Where we have defined
\begin{align}
\mathcal{U} & = \frac{1}{2} U^1 \tau^2 \\
\mathcal{T} & = (t^0+t^1),
\end{align}
which are characterising constants for the potential and hopping. Eq.~(\ref{eq:Constants1}) and (\ref{eq:Constants2}) agree with that of Ref.~\cite{Valiente2008}, where they consider a similar derivation in the case of a 1D Bose-Hubbard Hamiltonian.

We can now write out the approximate eigenfunction, using $z$, to be \cite{Valiente2008}
\begin{equation}
\begin{aligned}
\ket{\psi} \approx & \frac{\mathcal{N}}{\sqrt{(2^s s!)}} \sum_{z = -\frac{N+1}{2}}^{z = \frac{N+1}{2}} \Bigg[ H_s\left(\sqrt[4]{\frac{\mathcal{U}}{\mathcal{T}}} z\right) \\ &  \times e^{-\sqrt[4]{\frac{\mathcal{Y}}{\mathcal{T}}}\frac{z^2}{2}} (-1)^z \ket{z + \frac{N+1}{2}} \Bigg],
\end{aligned}
\label{eq:HarmonicWavefunctions} 
\end{equation}
and the eigenvalues to be
\begin{equation}
K_s \approx  E_{\mathrm{off}} - 2 \sqrt{\mathcal{U T}}\left(s + \frac{1}{2}\right) ,
\label{eq:HarmonicEnergy}.
\end{equation}
where $s = 0,1,...,N-1$, $H_s$ is the $s$th Hermite polynomial and $\mathcal{N}$ is a normalization constant. In Fig.~\ref{fig:WellApproximation} of the main text Eq.~(\ref{eq:HarmonicWavefunctions}) and Eq.~(\ref{eq:HarmonicEnergy}) give the analytical eigenfunctions and energies respectively.

\section{Double Filling Derivation}
\label{App:DoubleFilling}

As discussed in the main text, for $\nu = 2$ we have a similar form for the analogous tight-binding Hamiltonian to that of the SSH model \cite{Su1979,Su1980,Heeger1988,Asboth2016}. In this appendix, we focus on the solutions to the SSH Hamiltonian of Eq.~(\ref{eq:SSHHamiltonian}) in the main text. We can recast the SSH Hamiltonian into the form
\begin{equation}
\hat{H}_{SSH} = \sum_{j=1}^N (t - (-1)^j \delta t)\left[ \hat{b}_j^{\dagger} \hat{b}_{j+1} + \hat{b}_{j+1}^{\dagger} \hat{b}_{j} \right].
\label{eq:SSHAppendixHamiltonian}
\end{equation}
We now solve the stationary Schr\"{o}dinger equation for Hamiltonian~(\ref{eq:SSHAppendixHamiltonian}). The eigenfunction can be written as the sum over a set of coefficients at each site, $\ket{\Psi} = \sum_j \psi (j) \ket{j}$. Inserting the coefficient form of the eigenfunction into the stationary Schr\"{o}dinger equation we obtained a set of equations of the form
\begin{equation}
\begin{aligned}
(t - (-1)^j \delta t) \psi(j+1) \: + \: & (t - (-1)^{j-1} \delta t) \psi(j-1) \\ & = \epsilon \psi(j),
\end{aligned}
\label{eq:SingleEquations}
\end{equation}
where $\psi (j)$ will have the form $\psi(j) = \phi(j) e^{i k j}$, with $\phi(j)$ being the Bloch functions of the unit cell. We also know that the coefficients in the two site unit cell must only differ by a phase factor, therefore, we take an ansatz for the unit cell coefficients of the form
\begin{equation}
\phi(j) = \begin{pmatrix}
\pm e^{i \kappa} \\
1
\end{pmatrix},
\label{eq:UnitCellSSH}
\end{equation}
with the $\pm$ corresponding to the two bands of the model. Inserting the ansatz for the eigenfunction into Eq.~(\ref{eq:SingleEquations}) and solving for $\kappa$ and the quasi-energy $\epsilon$ we get
\begin{equation}
\kappa = \arctan \left[ \frac{\delta t}{t} \tan(k) \right]
\label{eq:PhaseFactor}
\end{equation}
and
\begin{equation}
\epsilon = \pm 2 \sqrt{t^2 \cos^2 (k) + \delta t^2 \sin^2 (k)}.
\label{eq:SSHQuasiEnergy}
\end{equation}
In the limit of $\delta t \rightarrow 0$, the expected solution of a homogeneous hopping model is recovered. 

The full set of eigenfunctions across the lattice can be found, after imposing the boundary conditions, to be of the form 
\begin{equation}
\Psi = \mathcal{N} \sum_j \left[ \phi_k(j) e^{i k j} - \phi_{-k}(j) e^{-i k j} \right] \ket{j},
\label{eq:SSHWavefunctions}
\end{equation}
where $\mathcal{N}$ is the normalisation coefficient. The quasi-momenta are
\begin{equation}
k = \frac{1}{N+1} \left( n \pi - \kappa(k) \right),
\label{eq:SSHMomentumSpectrum}
\end{equation}
with $n = 1,2,\dots,N/2$ (two bands). The $\kappa$ correction to the quasi-momenta is small, as $\delta t$ is small, and the allowed values of the quasi-momenta can be easily found numerically. The full energy spectrum for the double filling Hamiltonian~(\ref{eq:SingleParticleSSH}) is given by
\begin{equation}
K_n = 2t \mp \epsilon .
\label{eq:DoubleFillingAppendixEnergy}
\end{equation}


%
\end{document}